# Tube filling technique influence on superconductivity in $MgB_2$-based conductors prepared using laboratory made boron

Gianmarco Bovone[1,2], Davide Nardelli[3], Davide Matera[1,2], Maurizio Vignolo[2]

[1] Università degli studi di Genova, Physics Department (DIFI), Via Dodecaneso, 33 - 16146 Genova, Italy
[2] Istituto SPIN-CNR, Corso Perrone, 24 - 16152 Genova, Italy
[3] Columbus Superconductors SpA, Via delle Terre Rosse, 30 - 16133 Genova, Italy

Abstract – Here we report a systematic study of superconductive properties of mono-filamentary $MgB_2$-based wires, manufactured with different filling technique. A detailed comparison of the influence of filling technique and final heat treatment on superconducting properties has been given. Boron used was synthesized in laboratory following magnesiothermic reduction of boron oxide, purified thanks to several acid leaching and heat treated at high temperature, to enhance crystalline degree. Critical current density and critical temperature have been investigated at different synthesis/sintering temperature in order to evaluate their dependence to applied final heat treatment. Critical current density has been evaluated on short wire pieces by magnetic measurement in a MPMS 5.5T Quantum Design SQUID, while critical temperature has been measured with a four probe system, by the drop of resistivity during the cooling of the sample in a liquid helium dewar.

## 1. Introduction

The greatest challenge for superconductors is to manufacture conductors with higher transport current. Since the discovery of copper-based superconductors [1], researchers have been very active in this field with the developing of powder in tube (PIT) method [2]. Since the discovery of superconductivity in $MgB_2$ [3], this technique has been widely used to manufacture $MgB_2$ wires [4] and tapes [5]. PIT process is divided in *ex situ* [6,7,8] and *in situ* [9,10,11], that differs from the stage at which the reaction take place: for *ex situ* the tube is filled with pre-reacted powder, while for *in situ* it is filled with precursors of $MgB_2$.

Considering the reaction mechanism of $MgB_2$ [12], where molten Mg reacts with solid B, it is clear that B precursor is very important to manufacture a conductor with proper characteristics. B used in superconductive application is commercially available with different purity degree and average grain size. In general B is synthesized following magnesiothermic reaction optimized by Moissan [13,14]. Raw B has to be purified from boron oxide ($B_2O_3$) and MgO by several acid leaching but it is impossible to completely remove impurities. It is possible to synthesize B with different procedure, that permit to get a purer B, based on the reaction between $BCl_3$ and $H_2$ [15,16]. B synthesized in this way reaches purity over 99%, but with high production costs and risks.

Recently the importance of this material for the $MgB_2$ synthesis led some groups considering to synthesize B in their own laboratory, in order to tailor desired properties. Our research group developed a new procedure, based on the Moissan process, with the addition of a preliminary step [17]. In this way we are able to synthesize amorphous B with an average grain size between 60 to 120 nm [18], and a reduced grain size distribution. Furthermore it is possible to dope B precursor [19], making useless doping procedure on $MgB_2$ powder, e.g. ball milling. This is the main advantage on synthesizing B in laboratory, as shown by others groups [15,16]. In this work we evaluate $J_c$ and $T_c$ behavior of $MgB_2$ synthesized from pure laboratory-made B, and how different filling techniques and heat treatment can influence superconducting properties. In addition to typical *ex situ* and *in situ* technique, new ways for wire manufacturing have been developed, as an evolution of the previous ones. Table I reports a brief description about sample preparation.

## 2. Experimental Details

Ext and Int samples have been manufactured following classic *ex situ* and *in situ* PIT and results in wire cross section are those expected from literature data [20], as can be seen in figure 1. Stereomicroscope with high field depth has been used in order to evaluate the amount and dimension of void in transversal cross-section. $MgB_2$ for Ext sample has been synthesized at 920 °C for 20 min, from the reaction between $MgH_2$ and laboratory-made B. Int sample was fabricated by using an *in situ* process. Mg (Alfa Aesar, 98%) and Lab Synthesized B powders were used as starting materials with the composition of Mg:B = 1:2. The Mix technique, developed by Nardelli [21], is a based on the filling of the metal tube with a mixture of Mg and $MgB_4$, despite using B as precursor. This permit to reduce the volume reduction typical of the synthesis of $MgB_2$. $MgB_4$ was synthesized by direct reaction of magnesium turnings (Alfa Aesar, 98%) with laboratory-made B in an atomic ratio of 1:4. The mixture was pressed in a pellet, sealed by welding in an outgassed Ta crucible under pure Ar flow and then heat treated at 1100 °C for 5 days in a furnace. The resulting bulk was easily crushed in a blackish-violet powder. The $MgB_4$ powder was then mixed with Mg in a molecular ratio of 1:1 and subsequently the mixture has been used to fill the metal tube.

Table 1. Brief description of samples

| Wire sample | Process used | Mg Source | Final dimension [mm] | Metal sheet |
|---|---|---|---|---|
| **Ext** | *ex situ* [6] | $MgH_2$ powder | Square Section 0.97 x 0.97 | Nickel + Iron Barrier |
| **Int** | *in situ* [9] | Mg powder (Alfa Aesar, 98%) | Square Section 0.97 x 0.97 | Nickel + Iron Barrier |
| **Mix** | $MgB_2$ from $MgB_4$ [21] | Mg powder (Alfa Aesar, 98%) | Square Section 0.97 x 0.97 | Nickel + Iron Barrier |
| **Rli** | Reactive Liquid Infiltration [22] | Mg rod (99%) | Square Section 0.97 x 0.97 | Nickel + Iron Barrier |

A totally different approach has been used for the so-called "2$^{nd}$ generation $MgB_2$ wire", based on the Reactive Liquid Infiltration (Rli) technique developed by Giunchi[22]. In this method, Mg powder is substituted with a 3 mm in diameter Mg rod centred along the tube, then filled with B powder, weighted in order to maintain the correct stoichiometry. In Figure 1 are reported the transversal cross section of all the samples.

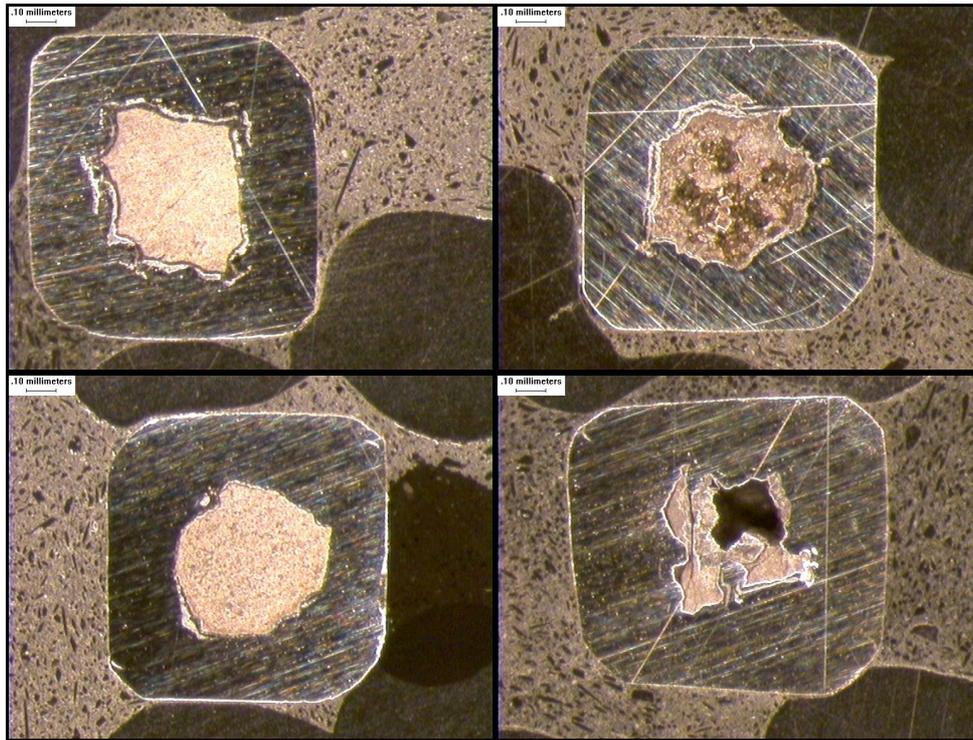

**Figure 1.** Trasversal cross section of MgB$_2$-based wires filled following ex-situ (a), in-situ (b), Mixed (c) and RLI (d) technique.

## 3. Results and discussion

Good magnetic $J_c$ values have been achieved as reported in figure 2, evaluated thanks to Bean Model [23] for wire sample. Measurements have been performed on short wire samples, cut from longer samples, heat-treated at different temperature for different times: 700 °C for 1 hour, 800 °C for 1 hour, 920 °C for 20 minutes and 1000 °C for 20 minutes. Different sintering/synthesis processes don't change significantly the trend of the curves: lowest $J_c$ values for Int wire and higher values at higher magnetic field for Rliwire. It is clear that Rli samples have highest $J_c$, but Mix sample shows lower heat treatment dependence. Rli sample heat treated in best conditions (700°C for 1 h) shows $J_c$ of $10^5$ A/cm$^2$ at 3 T, while the other samples (Ext, Int and Mix) reach the same value at least at 2 T. It is interesting the low dependence of $J_c$ to final heat treatment of Mix wire, not seen in previous work [21]. Worse performances of Int wire are related to low density packaging, due to voids in superconducting core, as can be seen in transversal cross section in figure 1. Exception is the sample heat treated at 800 °C for 1 hour that shows $J_c$ higher with respect *ex situ* wire. This can be due to a proper synthesized phase that could be not obtained with the other treatments. It should be noticed that B used in this work is different form commercial B [17], and it is possible that this kind of B has properties (morphology, grain-size, hardness) unsuitable for *in situ* PIT technique [24,25]. Good results have been achieved for Ext sample, manufactured following *ex situ* technique, in which our group have a great experience. The sample heat treated at 920°C for 20 minutes shows the highest $J_c$ at low field, reported for a pure MgB$_2$ *ex situ* wire manufactured from our research group [26].

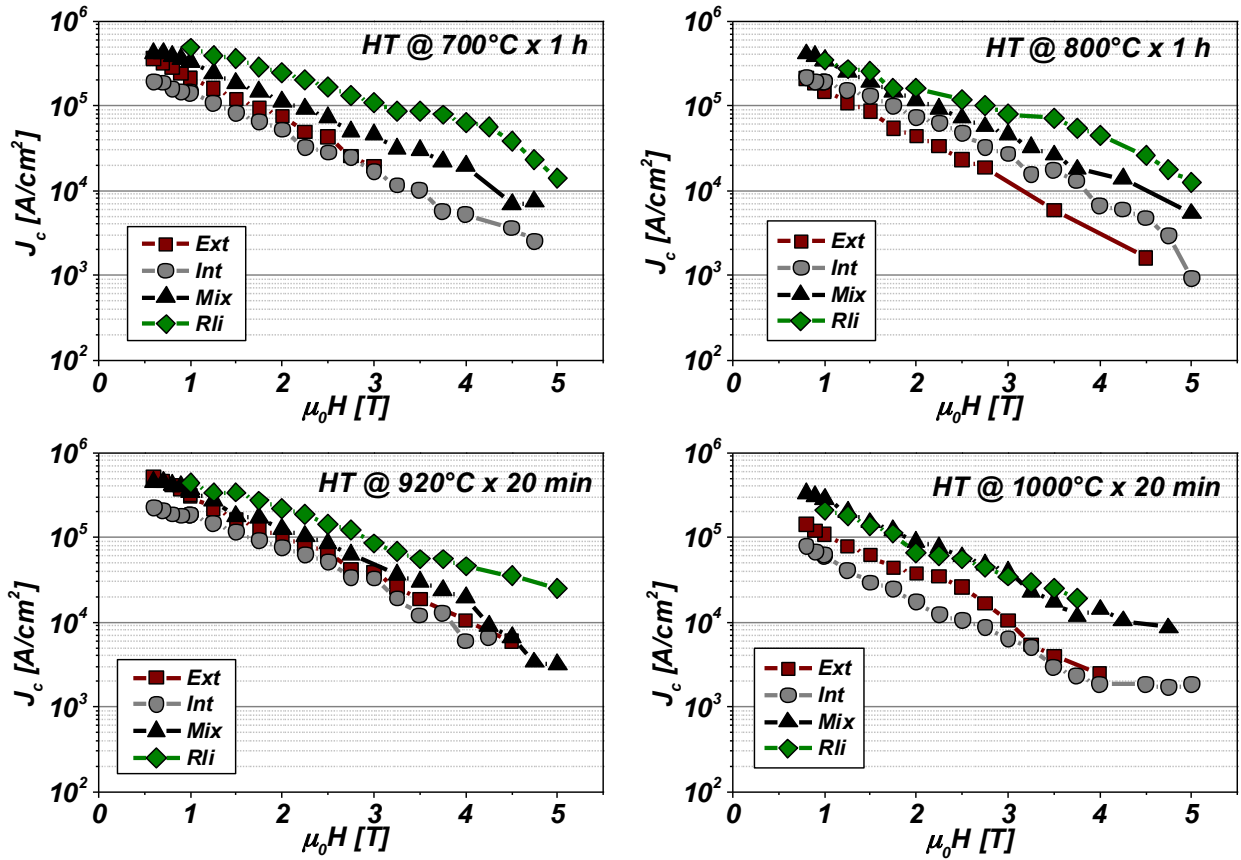

**Figure 2.** $J_c$ vs Field behaviour of short-piece wire sample heat treated at different temperature

Critical temperatures of wires are reported in figure 3. $T_c$ for Int sample are lower than the other samples. The only exemption, as well as for $J_c$, is the wire heat treated 1 h at 800 °C. Taking into account that heat treatment (for 1 h) at 800 °C is the best synthesis procedure in literature [25], this effect may be correlated to the use of very fine Mg powder, mixed with nanometric B. Lower grain size means higher superficial area, enhancing the surface oxidization of precursors then the MgO concentration in final $MgB_2$. MgO acts by lowering $T_c$ and $J_c$, as consequence of the reduced grain connectivity, as observed in this work and already reported in literature [26]. Rli and Ext samples have almost the same $T_c$ and the transitions are sharp due to good grain connectivity. For Ext wire the highest $T_c$ reported is for the sample heat treated at 920 °C for 20 min, as for $J_c$. As told before, this is the best heat treatment for ex-situ wire, and permits to reach best results in term of superconductive properties. Highest $T_c$s are reported for Mix sample, where synthesis process of $MgB_4$ seams to be able to "clean" B from impurities and this is confirmed from the higher $T_c$ (up to 38.5 K), despite using the same B precursor. Probably this "cleaning" effect is due to the migration of impurities at grain boundaries, where are less effective on $T_c$ lowering, and enhancing grain boundary pinning mechanism. Contrary to $J_c$s, the heat treatment at higher temperature (1000 °C for 20 min) does not effect on lowering $T_c$s, which remains similar to values comparable to the other heat treatments. This is due to the fact that $J_c$ is more affected by phase formation and grain connectivity, with respect $T_c$, and this can be achieved by a proper heat treatment.

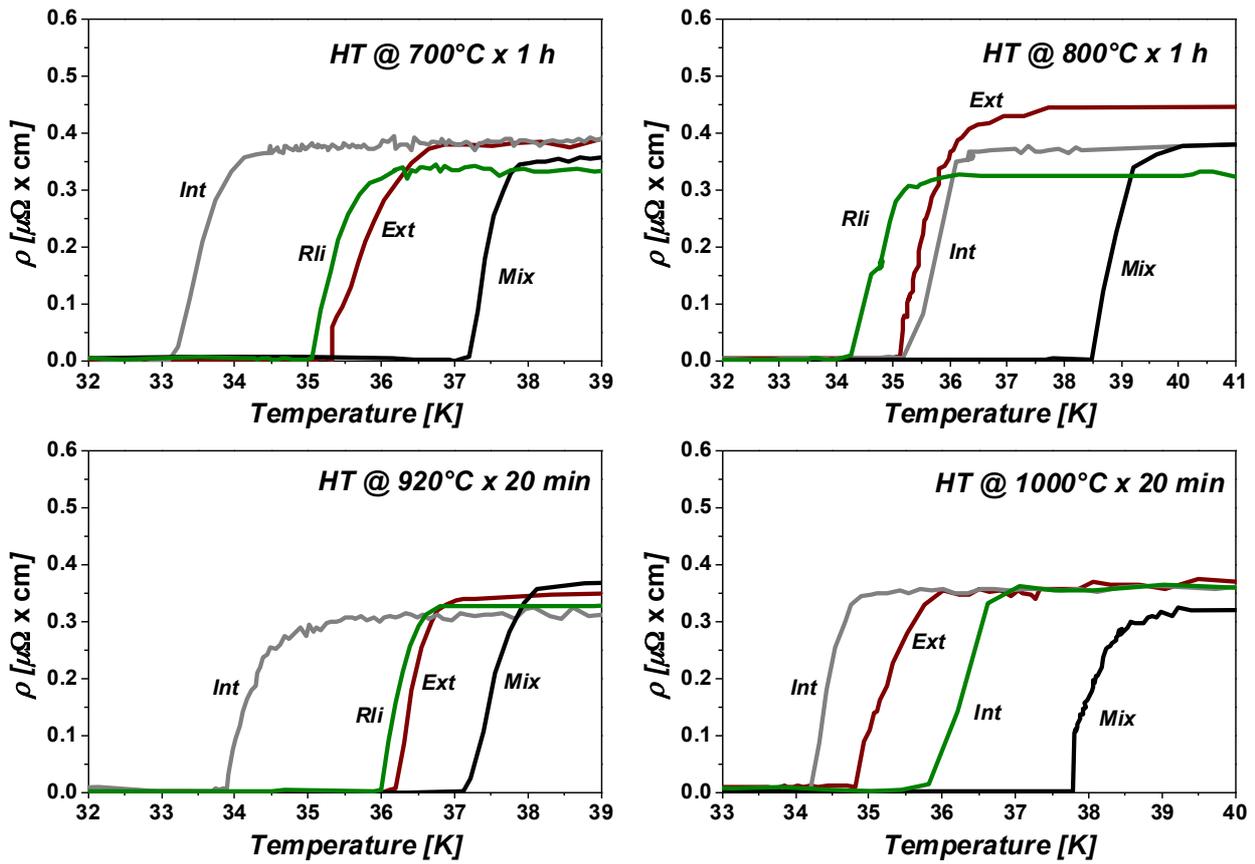
**Figure 3.** Resistive $T_c$ of wire sample heat treated at different temperature

In addition to the investigation with optical microscope, we analysed the transversal cross section with a Scanning Electron Microscope, equipped with an EDS probe for elemental analysis. From images reported in figure 4 it is clearly visible that the iron sheet broke during cold working process, and Ni from the tube diffused in $MgB_2$. This caused the formation of a thick reaction layer during heat treatment, which reduces the effective superconductive cross-section, besides reducing mechanical properties. Especially Ext and Int sample shows a thick reaction layer, due to the cracking of iron sheet in several points. This can partially explain lower performances, in terms of $J_c$, of these two samples. Rli sample shows no crack of iron sheet, and Ni diffusion in $MgB_2$ has not been observed. Moreover it has been observed a lack of Mg in Rli sample, that means $MgB_2$ phase is not completely synthesized with an high amount of unreacted B. Despite this effect high $J_c$ and $T_c$ have been reached.

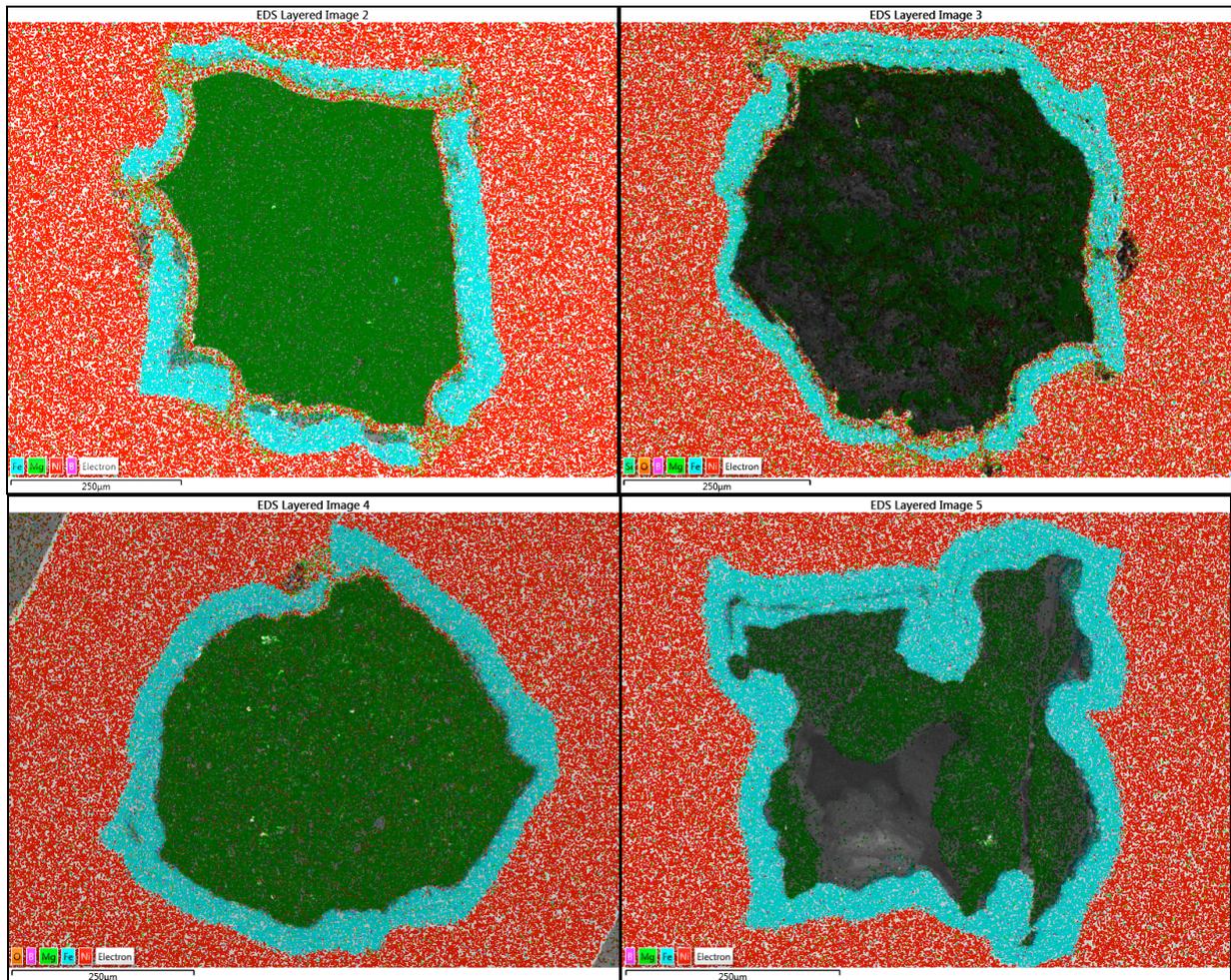
**Figure 4.** Elemental dispersion in transversal Cross-Section of the four samples, acquired with an EDS probe. Each colour is related to a specific element: Red for Ni, Cyan for Fe, Green for Mg and Violet for B.

## 4. Conclusions
Considering this results we can assume that B synthesized in laboratory, has proven to be an excellent precursor for superconductive application, especially in wire manufacturing. For 5 K application best results in terms of in-field behavior of $J_c$ are reached by filling tubes with Rli procedure, but with Mix technique $J_c$ is less affected by heat treatment and $T_c$ can be enhanced. Laboratory made B seems to be very suitable for Mix technique, and with the optimization of processing parameter performances can be enhanced, reaching Rli values. The main advantage of the match laboratory-made B and Mix technique is the ability of this process to reduce the effect of impurities in B. This effect can be noticed in the enhancing of $T_c$ and $J_c$, with respect to the classic *in situ* and *ex situ* process. Lower B grain-size had also the advantage to reach a very dense cross section of the wire, with low voids density. Unfortunately this effect was not observed for *in situ* wire, that shown a cross section with high voids density and low superconducting properties.